\documentclass[a4paper,10pt]{article}
\usepackage{graphicx}
\usepackage{amsmath}
\usepackage[utf8]{inputenc}
\usepackage{hyperref}
\usepackage{indentfirst}
\begin{document}
 
\begin{center}
\mbox{}

{\large{\bfseries  {Decision Making in Project Groups Formation: Students’ Perception and Reconciliation }}} 
\\[8pt]

\textbf{Rajeshwari K, Apurva Rohit Hegde, \\Drishika Patil}
\\[8pt]

\small{\textit{
Department of ISE, BMS College of Engineering \\
Bangalore, India \\
}}

*E-mail: {\tt rajeshwarik.ise@bmsce.ac.in, apurva.is18@bmsce.ac.in,\\ drishika.is18@bmsce.ac.in. }

\end{center}
\vspace{0.2cm}

\begin{abstract}
Academics is a huge repository of research avenue. Students tend to behave and adapt to the classroom based on their peer influences. Peers help in the increase of communication skills. Research shows group study is more effective than individual study. Group formation is influenced by several factors, like friends, demographic and linguistically similar. Toppers are more considerate but again based on comfort level friends are chosen. In this paper we analyze the change in preferences of group mates, and inferences are drawn over their reconciliations.\\
\textbf{Keywords} - Social Network Analysis, Preference Team, Preferred First Friend Next [PFFN], Preferred FCFS [PFCFS]
\end{abstract}
\vspace{10pt}

\section{Introduction}
Social Network Analysis (SNA) is the plotting and measuring of relationships, information flow between people, groups, and organizations. SNA provides both a visual and a mathematical analysis of relationships. Classroom dynamics is defined as the interaction between students and teachers in a classroom. The analysis of classroom dynamics helps in increasing the positivity of the classroom atmosphere, enhancing the student’s communication skills and making learning more comfortable. In this paper, Classroom dynamics is discussed to know how students want to interact with toppers or good performers of the class. Some students are known to be extroverts and having more friends, while some are introverts with lesser friends around. We make an observation on each student's opinions about their teammates’ performance and the student’s belief that a topper would help them learn more and improve their grades. An attempt is made to analyze student and peer relationships for a period of six months in Information Science and Engineering, BMS College of Engineering. Classroom peers are believed to impact learning by exchanging knowledge during discussions and project works. The effectiveness of this learning likely depends on classroom structure or team composition in terms of students’ capability. Peers can influence each other positively or negatively. Therefore, peer selection plays an important role in a student’s life. 

Data is collected in three phases for students of third semester in the course “Operating Systems”. As part of alternate assessments in this course, students were asked to prepare small projects in a team of three or four. Teams were framed by students themselves and there was no role play of the faculty of the course. Later during the semester, the admissions of lateral entrants were done after almost seven weeks of semester commencement. These students did not have any groups, the faculty on her own accord added individual lateral entrant students into already existing teams. As new students are not familiar to the already existing teams, this decision helped them to get into the team without any hurdles for them in choosing the teams. Stefanie Adriaensens et al., \cite{bib9} in a network perspective compares acceptance and rejection in the classroom interaction between Students Who Stutter (SWS) and their peers association in secondary education. Both SWS and their peers were considered by on an average six or seven classmates as ‘a friend’ and nominated as ‘no friend by one or two classmates’. The result obtained was that mostly both of the groups didn’t differ much in structural position in the class group (degree, closeness, betweenness). 
B Vasanthi et al., in paper \cite{bib5} has a dataset from a university in a remote village and another university in Mangalore city. The friendships and interpersonal communications vary from each of the universities. Reciprocity is found out for both the universities, which refers to responding to a positive action with another positive action to further maintain and strengthen social bonds. 
The aim of the work is to find out how the students' choices of partnerships in the teams vary, how they thrive to be with toppers, how they feel about working with friends. An analysis of changing of their teams, will that change really get implemented in the next coming semester group work was observed. A SNA perspective to the problem statement reveals the various friendships network characteristics, and the graph mutations based on the student choices.

\section{Literature Survey}
Toriumi Fujio et al. \cite{bib1}, aims at proving that grouping in a classroom community effectively decreases isolation and loneliness amongst students. Heider’s balance theory is used to analyze the network. Multi-agent simulation is used to analyze the teacher's influence on students. Sameep Mehta et al. \cite{bib2}, analyzes two algorithms to rank the interaction networks. Ranking Mechanisms for Graph Sparsification \& Ranking Mechanisms for Influence Maximization are used to determine key nodes in the network. Sampath Kameshwaran et al., examines in \cite{bib3} ranking techniques for agents of a Service Delivery setting. This is based on the agent's importance in the network structure and the value generated by his/her interactions. The ranking algorithm is based on extension of eigenvalue methods and it ranks the agents based on their importance and contribution in the value creation process. In \cite{bib4}, Qi Xuan et al., highlights the four important attributes of social synchrony i.e. Depth of action, Breadth of impact, Heterogeneity of role, Emergence of phenomenon. Keval Vora et al. analyzes in paper \cite{bib6} about evolving graphs or mutating graphs (graphs that change over time). The graph can change by addition or elimination of edges and vertices. This paper analyses graphs using two methods: Fetch Amortization and Process Amortization. Processing of evolving graphs is accelerated using both the two optimizations. Fetch Amortization reduces remote fetches while Processing Amortization accelerates termination of iterative algorithms. M. Broom et al., in paper \cite{bib7} talks about evolutionary dynamic models. For a regular graph, the fixation probability does not affect the commencement point. In the star graph, the mutant does signiﬁcantly better when it starts on the boundary and similarly, starting at the end of the line guarantees the highest ﬁxation probability on the line. Unwell-mixed graphs have more routes to ﬁxation and the contest is resolved far more quickly than on tree-like structures, especially the star, and the difference can be several orders of magnitude even for very small graphs. Aynaz Taheri et al., gives the reasoning in paper \cite{bib8} about a sequence-to-sequence framework which is used to train the encoder-decoder model. Dynamic graphs are used for a troop of GPS-tracked baboons living in the wild in Mpala Research Centre, Kenya. The goal is to learn graph embeddings such that graphs having similar topological structure and temporal dynamics lie close to one another in the embedding space. Baomin Xu, in the paper \cite{bib12} discusses an algorithm of job scheduling in cloud computing based on the Berger model. In the job scheduling process, the algorithm establishes dual fairness constraints. The first constraint is to classify user tasks by QoS preferences, and establish the general expectation function in accordance with the classification of tasks to restrain the fairness of the resources in the selection process. In cloud computing, QoS is a measure of user satisfaction with cloud services. The QoS parameters considered in this paper are completion time and Bandwidth. The second constraint is to define resource fairness justice function to judge the fairness of the resources allocation. The proposed algorithm in this paper provides an effective implementation of user tasks, and with better fairness. R. Lars Backstrom et al., in paper \cite{bib14} uses decision trees to compare the reasons for the formation of growth of groups in different communities. The reasons considered are the number of friends already in the community as well as the network formed within the friends already in the community. The paper also describes the reasons for community growth.

\section{Methodology}
\subsection*{Dataset description}
Dataset was prepared for students of III semester, Information Science and Engineering, BMS College of Engineering for the AY 2019-20. As part of an assessment, the students were required to submit a project for the course Operating Systems within a period of four months. They were allowed to group themselves in teams of four. The dataset is collected in three phases. In phase 1, we collected the students’ groups information from the faculty coordinator of the course and we will term this phase as “OS Team”. 

As shown in Table~\ref{table1}, the total number of students are n = 57. This includes regular students, diploma (lateral entrants), backlog and branch change students. 
\begin{table}[tbh]
\caption{\label{table1} Student Details in the Dataset.}
\vspace{2mm}
\centering
  \includegraphics[width=0.3\textwidth]{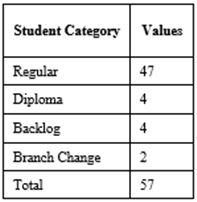}
\end{table}

Georgios Kanakis, in paper \cite{bib11} discusses an essential and flexible framework to allow engineers to form collaborating groups the way they decide. Collaborations are of three types synchronous (instant), asynchronous (triggered) and mixed collaboration. This paper aims to allow engineers collaborating instantly or with triggered mode allowing them to apply their individual work style or the style that their role in the project requires. The framework ensures that these conditions are held within the group by formally verifying the collaboration language and its change propagation algorithm using Alloy.  

Further after the assessment of the course, at the end of the semester, students were asked a couple of questionnaires to analyze how they felt about their previous group’s performance. A google form was shared to collect their opinions on the following questionnaire.\\ 
-	Teammates rating (P1) \\
-	Open communication among teammates (P2)\\
-	Information sharing among teammates (P3)\\
-	Shared decision making among teammates (P4)\\
-	Skill sharing with teammates (P5)\\
-	Sharing of knowledge with teammates (P6)
\begin{figure}[!h]
  \centering
  \includegraphics[width=0.6\textwidth]{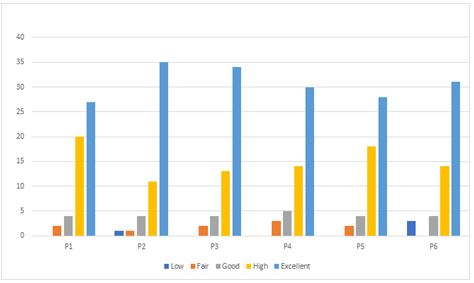}
  \caption{ Questionnaire Responses\label{fig:1}}
\end{figure}\\

We observed from the numericals, that many of the students were not pleased in the groups of Operating System course. There are many reasons behind this dissatisfaction. During the first year of the course, students were spread across different sections. Then later they were rearranged into two sections during third semester. Familiarity amongst the students was lesser and they were hesitant to make groups. Another reason could be, as the diploma students were assigned into a group randomly by the teacher, they did not suit the group, also these students joined the groups midway through the project work. From Fig.\,\ref{fig:1}, it is observed that P2 values for both low and fair rating were 1. Less knowledge was transferred to diploma students, as time would have been a concern. P6 values for low rating is 3, given for the last questionnaire i.e. sharing of knowledge with teammates (P6). The reason could be that the students who were weak in grasping knowledge, involved in the given project, were excluded from participation in the development of the project and hence felt left out and thus these set of students desired for a new group. Keeping this in mind we collected students' choices of preferred teammates. This is phase 2 dataset. Sancheng Peng \cite{bib10}, aims to provide a broad research guideline of existing and ongoing efforts via social influence analysis in large-scale social networks, and to help researchers better understand the existing work, and design new algorithms and methods for social influence analysis. The influence maximization algorithm is designed to find the most influential top-k nodes. In our dataset, we have the opinions of students for a successful team to be framed, we termed this as “Preference Team”. This list is a directed graph. Students were asked three people whom they prefer to be teamed for the next project work in the upcoming semester. A survey was conducted to know if students would benefit from having a topper/good programmer in the team. The survey concluded that 14 students consider toppers/good programmers essential for improvement of themselves or the team, whereas 9 students don't find them important, 30 students were not sure if the toppers would influence them and were confident about their individual work. 

Phase 3 dataset constitutes the teams which were formed in the IV semester for the same set of students for the course “Seminar Project Work”. Using this information, we check if the preference team has been reflected in this team formation. Phase 3 consists of seminar teams formed which are referred to as “New Team”. This dataset was again shared by the faculty coordinator. 

\begin{table}[tbh]
\caption{\label{table2} Statistics of Friendship Network.}
\vspace{2mm}
\centering
  \includegraphics[width=0.6\textwidth]{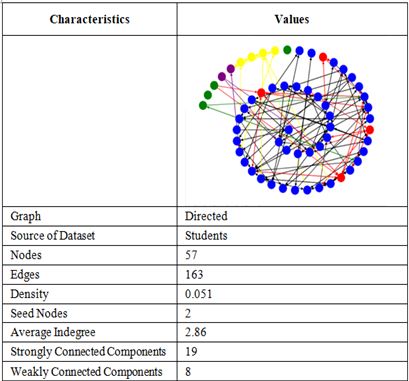}
\end{table}

We also prepared a Friendship Network which consisted of information about the friends of all the students present in the class. The number of friends per student varies from 0 to 4. The statistics of the network are visible in Table~\ref{table2}. It is perceived from Table~\ref{table2} that the density is low. Low density represents the lack of strong friendship amongst the students of the class. The reasons for this could be because of the lack of familiarity amongst the students since they belonged to different sections in first year. Another reason could be that in many cases friendship is one-sided. Based on indegrees, we found students with the highest indegree were nodes 24, 39, 40. The node 39 is the III sem topper and the nodes 24 and 40 are friendly and are extroverts. This shows that friendship is forged between individuals based on their personalities and not only marks.  
\begin{table}[tbh]
\caption{\label{table3} Topper Nodes for Previous Semesters.}
\vspace{2mm}
\centering
  \includegraphics[width=0.9\textwidth]{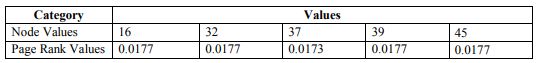}
\end{table}

The Topper Nodes are the students who have the highest marks in the I, II and III semesters for the Academic year 2018-2020, as shown in Table~\ref{table3}. The number of students who preferred toppers were 26. The number of students who got toppers when the seminar teams were formed i.e. in the new list are only 8. This shows that the desired team is not formed due to the unavailability of the preferred teammates. 
\subsection*{Evaluation of Dataset}
In the preference list, we found seed nodes, notably a single student was found who was not willing to participate in a team, similarly the same student was not preferred in the preference list by his classmates. Reasons for this are very speculative, as the student was good at project work, and is neither an introvert.

Furthermore, we queried the following for the preference team list:\\
-	Does a good performer choose another good
performer?\\
-	Does the preference team consist only of toppers?\\
-	Does the preference team consist only of friends?\\
-	Whom do the diploma and branch change students prefer? \\

Nodes with the highest indegree in preference team are -\\ 
Node 16 with Indegree 3\\
Node 32 with Indegree 3\\
Node 22 with Indegree 2\\
The nodes 16 and 32 have been toppers in the 1st and 2nd semester and thus have a greater preference among the diploma and branch change students. 

We made a comparison between the grades of a student and that of their preferred choice of students to justify the students’ choice. The preference of the pupil can have three different alternatives - one is that a student chose someone who is academically better than themselves in studies, the other choice is that the student is a weaker performer than themselves in studies and the last possibility is that the student is equivalent to them in studies. We made this compatibility study by utilizing the flag values. 
\begin{figure}[!h]
  \centering
  \includegraphics[width=0.9\textwidth]{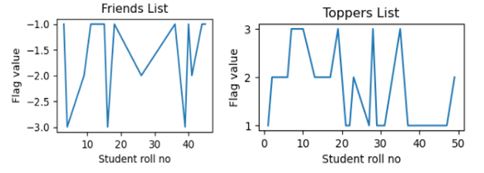}
  \caption{ Plots for friends over toppers\label{fig:2}}
\end{figure}

Fig.\,\ref{fig:2} shows these plots of selections of friends over the toppers. The flag value keeps account of whether a student forms a team with students who are better in academics. The negative flag value indicates that a student forms a team with students who are relatively less good in academics. This means that students prefer to form teams with their friends rather than toppers or students who are relatively better at academics. The flag values for positive, negative and neutral are 22, 14 and 21 respectively. This brings us to an important conclusion that students want to progress hence they tend to choose students who are academically as good or better than them.

\subsection*{Algorithms Implementation}
We consider the following variables, for n = 57.\\
Let V = \{v\textsubscript{1}, v\textsubscript{2} …. v\textsubscript{n}\},represent all the set of nodes in the graph, where v\textsubscript{i} represents the i\textsuperscript{th} node in the set.

Let F = \{v\textsubscript{1}: F(V\textsubscript{1}), v\textsubscript{2}: F(V\textsubscript{2}), .... v\textsubscript{n}:F(V\textsubscript{n})\}, be the dictionary of friends for the vertex v\textsubscript{i}, where F(V\textsubscript{i}) represents the set of friends for the vertex v\textsubscript{i}.

\centerline{F(V\textsubscript{i}) = \{v\textsubscript{j}\}, be the set of m friends for the node v\textsubscript{i}}.

Let P = \{v\textsubscript{1}: P(V\textsubscript{1}), v\textsubscript{2}: P(V\textsubscript{2}), ....., v\textsubscript{n}: P(V\textsubscript{n})\}, be a dictionary of preferred partners and P(V\textsubscript{i}), be the set of preferred partners of the vertex v\textsubscript{i}

\centerline{P(V\textsubscript{i}) = \{v\textsubscript{j}\}, be the set of m preferred partners for the node v\textsubscript{i}.}

Let 'Allot' be a dictionary to hold all the team allotment details for each iteration of the two algorithms.

The aim of the algorithms is to allot students into teams of four. We chose four per team as per faculty coordinator’s instruction. The students’ friends and preferred teammates data is considered.
Satisfaction Index [SI] is populated to get the number of teams which are completely complacent i.e. the number of teams which have members of four in it. Seed Nodes [SN] are isolated nodes who are hesitant to be a part of any team and thus form a team with only themselves in it. At the end of the algorithms, we find the SI and SN to know the success of iterations. Kulkarni \cite{bib15} discusses various scheduling algorithms to know the interplay of the nodes selections and their successions.

\subsection*{Algorithm 1:  Preferred First Friend Next [PFFN]}
The team allotment is done by considering First Come First Serve (FCFS) Model. In the first iteration, the first USN is given his/her preferred team list, provided the same set of students prefer him as the teammate. If not the first USN would be allotted his friends as his teammates. The allotted students are unavailable for the next set of students who would prefer them. In the same manner we pick the next USN, check if these students were allotted previously, else allot and prepare the team. In the second iteration the allotment starts from the second USN and the allotment continues in a similar fashion. Satisfaction index and seed node are stored for each iteration Kanniga Devi et al., in paper \cite{bib13} puts forward a new algorithm i.e., independent, dynamic, batch mode algorithm and compares it with the other batch algorithms and concludes with the help of different graphs that it is a viable algorithm. \\
1: Start\\
2: Initialize the sets V, F, and P and x →0\\
3: Set i → x\\
4: if for any node v\textsubscript{i} there exists a node v\textsubscript{j} in P(V\textsubscript{i}):\\
\indent if for node v\textsubscript{j} the node v\textsubscript{i} exists is in P(v\textsubscript{j}):\\
\indent \indent then,\\
\indent \indent \indent Allot [v\textsubscript{i}] → v\textsubscript{j}\\
\indent \indent \indent Allot [v\textsubscript{j}] → v\textsubscript{i}
// friendship values are iterated for unavailable students\\
else if for v\textsubscript{i} if there exists a node v\textsubscript{k} in F(V\textsubscript{i}) and v\textsubscript{k} is not yet allotted to any other node:\\	\indent Then,\\
\indent \indent Allot [v\textsubscript{i}] → v\textsubscript{k}\\
\indent \indent Allot [v\textsubscript{k}] → v\textsubscript{i}\\
5: Repeat 4 for x \textless= i \textless total no of nodes and then for 0 \textless= i \textless x if it exists.\\
6: Store the SI and SNs for the iteration.\\
7: Repeat 3 to 6 for x \textless total no of nodes.\\
8: Stop\\

The time complexity of the algorithm is estimated to O (n\textsuperscript{2} (f+p)), where f is the length of the friends’ list and the p is the length of the preferred team for the node v\textsubscript{i}. 

\subsection*{Algorithm 2: Preferred FCFS [PFCFS]}
The team allotment is done by considering First Come First Serve (FCFS) Model. In the first iteration, the first USN is given his/her preferred team list. A team of size four is prepared. The allotted students are unavailable for the next set of students who would prefer them. In the same manner we pick the next USN, check if the preferences of these students are not allotted for previous USN, else allot and prepare the team. In the first iteration, the first USN gets the preference over the rest of the class students. In the second iteration the allotment starts from the second USN and the allotment continues in a similar fashion. In nth iteration, the last USN, i.e. backlog student gets his/her preference first over the entire class. Thereby his/her choice also gets executed once, and impacts the entire class allotment.  Every iteration, SI and SN are stored. An iteration which yields, highest value of SI and lowest values of SN can be considered as the best iteration and also a starting point for team allotment. \\
1: Start\\
2: Initialize the sets V, P and set x→0\\
3: Set i → x\\
4: If for any node v\textsubscript{i} there exists a node v\textsubscript{j} in P(V\textsubscript{i}) which is not yet allotted to any other node:\\
\indent Then,\\
\indent \indent Allot [v\textsubscript{i}] → v\textsubscript{j}\\
\indent \indent Allot [v\textsubscript{j}] → v\textsubscript{i}\\
5: Repeat 4 for x \textless= i \textless total no of nodes and then \\
\indent for 0 \textless= i \textless x if it exists.\\
6: Get the satisfaction index for the iteration.\\
7: Repeat  3 to 6 for x \textless total no of nodes.\\
8: Stop\\

The time complexity of this algorithm is estimated to O(n\textsuperscript{2}(p)), where p is the length of the preferred team for the node v\textsubscript{i}. 

\subsection*{Comparison of PFFN and PFCFS}
\begin{table}[tbh]
\caption{\label{table4} SI and SN for the Toppers as per PFFN and PFCFS algorithms.}
\vspace{2mm}
\centering
  \includegraphics[width=0.6\textwidth]{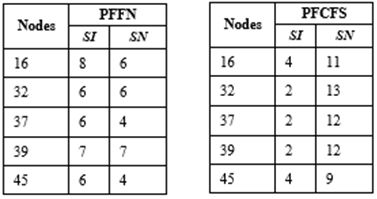}
\end{table}
Table~\ref{table4} shows the results for SI and SN, and the iteration considers the topper nodes to be starting points. The iteration which provides the highest value of SI and lowest value of SN in both the algorithms is considered as an ideal case for team allotment. Since it reveals that the team thus formed will have the least number of non-participating students i.e., least number of seed nodes and has the most satisfied crowd of students with highest SI.\\

\begin{table}[htb]
\caption{\label{table5} Iteration statistics with the highest similarity with the new list.}
\vspace{2mm}
\centering
  \includegraphics[width=0.9\textwidth]{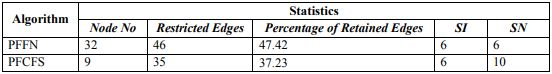}
\end{table}

Table~\ref{table5} shows the comparison between two algorithms for the maximum number of retained edges with the new list. It shows the SI and SN to be 6 and 10 for each algorithm respectively.\\

\begin{table}[htb]
\caption{\label{table6} Highest SI iteration.}
\vspace{2mm}
\centering
  \includegraphics[width=0.9\textwidth]{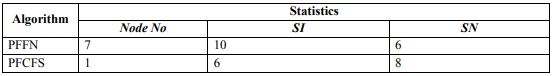}
\end{table}

Table~\ref{table6} shows the results of the algorithms for highest values of SI.

\section{Results and Discussions}
The analysis made on the queries given in the evaluation of dataset were concluded as given below:\\
-	A good performer does choose another good performer, since out of the 5 toppers of the class, 3 preferred another topper as their preference for teammate.\\
-	The majority of students do not choose toppers as their preferences. We found 17 students chose toppers.\\
-	The preference team mostly consists of friends because 44 out of 57 students chose only friends as their preferences.\\
-	It is found that diploma and branch change students prefer to have a topper in their team. Amongst 6 branch change and diploma students 4 chose toppers, as they lack familiarity with the students and the course.\\

Table~\ref{table7} shows the characteristics of the various networks considered. OS team and New list is an undirected graph, whereas the preference team is a directed graph, similar to friendship network. Students who didn't share their preference list and did their projects alone are seed nodes. These students are backlog students in most cases (i.e. nodes 48, 55, 56). The backlog students don't attend all the classes and spend less class contact hours with the other students therefore they are not an active part of team formation. Some students who are seed nodes and not backlog students (i.e. node 34) are students who are not willing to socialize and prefer their own company over others.

Most influential nodes in preferred teams were found using pagerank algorithm. The salient nodes were: Topper of first year (37), toppers of III semester (39) and the influential student/Friendly student (31). 
Therefore, it is observed that the most important nodes are the toppers, influencers and students with a big friends group.

\begin{table}[tbh]
\caption{\label{table7} Statistics of networks.}
\vspace{2mm}
\centering
  \includegraphics[width=0.9\textwidth]{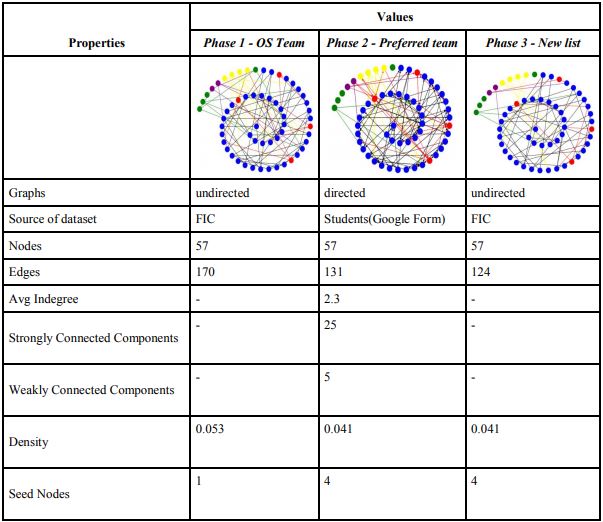}
\end{table}

\section{Conclusions}
A classroom is a best example for the homophily to be studied on. Peer learning contributes more to the overall development of the student than an individual learning. An attempt is made in this paper to learn the student's conceptions of building social links within the classroom. Every student has a wish list of friends, and confidante, and teams to work with. Due to various influences across the classroom, students tend to be satisfied with the friends whom they know better than an unknown classmate. The students tend to reconcile with previous teammates, due to non-availability of their fancied list. We found a satisfaction index, when a particular student’s choice is granted to see the status of the rest of the class. PFFN and PFCFS are being discussed. Also, as a future scope, we can look into some algorithms which can help us to allot the teams and draw more satisfaction from the teammates. 


\small{
}

\end{document}